\documentclass[twocolumn,pre,superscriptaddress,floatfix]{revtex4-1}

\usepackage{graphicx}
\usepackage{amsmath}
\usepackage{amssymb}
\usepackage{bm}
\usepackage{times}

\usepackage[pdftex]{hyperref}
\hypersetup{colorlinks=true,linkcolor=blue,citecolor=blue,urlcolor=blue}

\newcommand{\pbc}{\pi}
\newcommand{\abc}{\overline{\pi}}

\begin{document}

\title{Fractal dimension of interfaces in Edwards-Anderson spin
glasses for up to six space dimensions}

\author{Wenlong Wang}
\email{wenlongcmp@gmail.com}
\affiliation{Department of Theoretical Physics, Royal Institute of Technology, Stockholm, SE-106 91, Sweden}
\affiliation{Department of Physics and Astronomy, Texas A\&M University,
College Station, Texas 77843-4242, USA}

\author{M.~A.~Moore}
\affiliation{School  of
Physics and Astronomy, University of Manchester, Manchester M13 9PL, UK}

\author{Helmut G.~Katzgraber}
\affiliation{Department of Physics and Astronomy, Texas A\&M University,
College Station, Texas 77843-4242, USA}
\affiliation{1QB Information Technologies (1QBit), Vancouver, British
Columbia, Canada V6B 4W4}
\affiliation{Santa Fe Institute, 1399 Hyde Park Road, Santa Fe,
New Mexico 87501, USA}

\date{\today}

\begin{abstract}

The fractal dimension of domain walls produced by changing the boundary
conditions from periodic to anti-periodic in one spatial direction is
studied using both the strong-disorder renormalization group and the
greedy algorithm for the Edwards-Anderson Ising spin-glass model for up
to six space dimensions.  We find that  for five or less space
dimensions, the fractal dimension is less  than the space dimension.
This means that interfaces are not space filling, thus implying replica
symmetry breaking is absent in  space dimensions fewer than six.
However, the fractal dimension approaches the space dimension in six
dimensions, indicating that replica symmetry breaking occurs above six
dimensions. In two space dimensions, the strong-disorder renormalization
group results for the fractal dimension are in good agreement with
essentially exact numerical results, but the small difference is
significant. We discuss the origin of this close agreement. For the
greedy algorithm there is analytical expectation that the fractal
dimension is equal to the space dimension in six dimensions and our
numerical results are consistent with this expectation.

\end{abstract}

\maketitle

\section{Introduction}
\label{sec:intro}

One of the outstanding problems of statistical physics is the nature of
the ordered phase of spin glasses. While this problem is primarily of
interest to researchers in statistical and condensed matter physics,
spin-offs from its study have found their way into different fields of
research, such as computer science and neural networks. Unfortunately,
standard methods used in condensed matter physics, such as the
renormalization group and mean-field theory, have resulted in a confusing
situation for the nature of the spin-glass state. The picture that
derives from mean-field theory---valid for infinite-dimensional
systems---is that of replica symmetry breaking (RSB)
\cite{parisi:79,parisi:83,rammal:86,mezard:87,parisi:08}.  However,
results using real-space renormalization group (RG) methods---which are
better for low-dimensional systems---suggest a spin-glass state with
replica symmetry \cite{moore:98,
monthus:15,wang:17a,angelini:15,angelini:17}.  The purpose of this work
is to present additional numerical results beyond those presented in
Ref.~\cite{wang:17a} that suggest that in space dimension $d \le 6$ the
low-temperature phase of spin glasses is replica symmetric, and that it
is only for dimensions $d > 6$ that RSB prevails.

In the absence of RSB, the droplet picture (DP)
\cite{mcmillan:84a,bray:86,fisher:88} is expected, i.e., when $d \le 6$.
In the DP the low-temperature phase is replica symmetric and there is no
de Almeida-Thouless line \cite{dealmeida:78} in the presence of an
applied field. Its properties are determined by the excitation of
droplets whose free-energy cost on a length scale $\ell$ goes as
$\ell^{\theta}$ and which have fractal dimension $d_{\rm s} < d$. In the
RSB picture there exist system-size excitations which have a free-energy
cost of $O(1)$ and which are space filling, i.e., have $d_{\rm s}=d$.
Thus by investigating the value of $d_{\rm s}$ of interfaces in the
low-temperature phase, it is possible to determine whether the
low-temperature state is best described by RSB or DP.  Direct Monte
Carlo simulations to determine the value of $d_{\rm s}$ in $d = 3$ have
proved inconclusive (see, for example, Ref.~\cite{katzgraber:01} and
references therein). This is because the numerically accessible system
sizes in equilibrated simulations are just too small to distinguish RSB \cite{marinari:00, billoire:12a}
from DP behavior \cite{wang:17c}. One advantage of using real-space RG
methods such as the strong-disorder renormalization group (SDRG) method
is that one can study much larger system sizes than can be thermalized
in Monte Carlo simulations. Therefore, in this study we use SDRG, as
well as a greedy algorithm to estimate $d_{\rm s}$ for spin glasses in
different space dimensions $d$.

The paper is structured as follows. In Sec.~\ref{sec:formalism} we
introduce the model studied, and describe how by studying the link
overlap one can determine the fractal dimension of interfaces. In
Sec.~\ref{sec:SDRG} we give some details of the SDRG procedure as
developed by Monthus \cite{monthus:15} and outline why it is expected to
work better in two dimensions than in six space dimensions. Our results
for $d_{\rm s}$ in dimensions $d=2$, $3$, $4$, $5$, and $6$ are reported
in Sec.~\ref{sec:SDRGresults}. The greedy algorithm (GA) used here as
well is described in Sec.~\ref{sec:greedy}.  We conclude with a brief
discussion in Sec.~\ref{sec:discussion}.

\section{Model and observables}
\label{sec:formalism}

We study the Edwards-Anderson (EA) Ising spin-glass model
\cite{edwards:75} on a $d$-dimensional hypercubic lattice of linear
extent $L$ described by the Hamiltonian
\begin{equation} 
H = - \sum_{\langle ij \rangle} J_{ij} S_i S_j,
\label{eq:ham} 
\end{equation} 
where the summation is over nearest-neighbor bonds and the random
couplings $J_{ij}$ are  chosen from a standard Gaussian distribution
of unit variance and zero mean. The Ising spins take the values $S_i
\in \{\pm 1\}$ with $i = 1,2, \ldots, L^d$. 

The fractal dimension $d_{\rm s}$ can be obtained from the link overlap 
\begin{equation}
q_{\ell} =\frac{1}{N_b} 
\sum_{\langle ij \rangle} 
S_i^{(\pbc)}S_j^{(\pbc)} 
S_i^{(\abc)}S_j^{(\abc)} 
\left(2 \delta_{J_{ij}^{\pbc},J_{ij}^{\abc}} - 1\right).
\label{eqn:pqdef}
\end{equation}
Here $S_i^{(\pi)}$ and $S_i^{(\abc)}$ 
denote the ground states
found with periodic $(\pi)$ and antiperiodic $(\abc)$ boundary conditions, 
respectively. One can change from periodic to antiperiodic
boundary conditions by flipping the sign of the bonds crossing a
hyperplane of the lattice. $N_b$ is the number of nearest-neighbor bonds
in the lattice which for a $d$-dimensional hypercube is given by $N_b=d
L^d$. The $L$ dependence of the quantity  $\Gamma$  determines $d_{\rm s}$ via
\begin{equation}
\Gamma \equiv 1-q_{\ell}=\frac{2\Sigma^{\rm DW}}{d L^d} \sim L^{d_{\rm s}-d},
\label{eqn:gammadef}
\end{equation}
where $\Sigma^{DW}$ is the number of bonds crossed by the domain wall bounding the flipped spins \cite{hartmann:02}.
 The domain wall could be fractal, i.e., its
``length'' $\Sigma^{DW} \sim A L^{d_{\rm s}}$. If the interface were
straight across the system, its length would be $\sim L^{d-1}$. In the
RSB phase $d_{\rm s}=d$, so that $d-1 \le d_{\rm s} \le d$. The SDRG
(and also the GA) methods are just means by which one can determine the
(approximate) ground states  needed in Eqs.~\eqref{eqn:pqdef} and
\eqref{eqn:gammadef}.

\section{The SDRG algorithm}
\label{sec:SDRG}

In this Section we outline the SDRG method as described by Monthus in 
Ref.~\cite{monthus:15}. For each spin $S_i$, the local field is 
\begin{eqnarray}
h^{\rm loc}_i && = \sum_{j} J_{ij}  S_j.
\label{hloci}
\end{eqnarray} 
The SDRG focuses on the largest term in absolute value in the sum corresponding 
to some index $j_{\rm max}(i)$
\begin{eqnarray}
 \max_{j} (\vert J_{ij} \vert) \equiv \vert J_{i,j_{\rm max}(i)}  \vert.
\label{omegai}
\end{eqnarray}
The question for the accuracy of the SDRG is whether the local field 
$h^{\rm loc}_i$
\begin{eqnarray}
h^{\rm loc}_i && =  
J_{i,j_{\rm max}(i)} S_{j_{\rm max}(i)}+ \sum_{j \ne j_{\rm max}(i)}  J_{ij} S_j
\label{hloci2}
\end{eqnarray}
is dominated by the first term. 

The ``worst case'' is  when the spins $S_j$ of the second term in
Eq.~(\ref{hloci2}) are such that $( J_{ij} S_j)$ all have the same sign;
their contribution to the local field is then  maximal. Monthus
introduced the difference
\begin{eqnarray}
\Delta_i && 	\equiv \vert J_{i,j_{\rm max}(i)}  \vert 
		- \sum_{j \ne j_{\rm max}(i)}  \vert J_{ij}  \vert.
\label{deltai}
\end{eqnarray}
For $\Delta_{i_0}>0$, the sign of the local field $h^{\rm loc}_{i_0}$
is determined by the sign of the first term $J_{i_0j_{\rm
max}(i_0)} S_{j_{\rm max}(i_0)} $ for all values taken by the other
spins $S_j$ with $j \ne j_{\rm max}(i_0)$;
\begin{eqnarray}
{\rm sgn} ( h^{\rm loc}_{i_0} ) && 
=S_{j_{\rm max}({i_0})} {\rm sgn} \left[ J_{{i_0},j_{\rm max}({i_0})}\right].
\label{hlocisgn}
\end{eqnarray}
Then the spin $S_{i_0}$ can be eliminated via
\begin{eqnarray}
S_{{i_0}} =  S_{j_{\rm max}({i_0})} {\rm sgn} \left[J_{{i_0} j_{\rm max}({i_0})}\right]
\label{elimsi0}
\end{eqnarray}
so that Eq.~\eqref{eq:ham} becomes
\begin{eqnarray}
H &=&  	-\vert J_{{i_0}j_{\rm max}({i_0})}\vert 
	- \sum_{(i,j)\ne i_0}J_{ij}^{\rm R} S_i S_j,
\label{hsgdeci}
\end{eqnarray}
where the renormalized couplings connected to the spin $S_{j_{\rm max}(i_0)}$ are 
\begin{eqnarray}
J^{\rm R}_{j_{\rm max}(i_0),j} = 
J_{j_{\rm max}(i_0),j}+ J_{i_0,j}  {\rm sgn} \left[J_{i_0 j_{\rm
max}(i_0)}\right].
\label{jr}
\end{eqnarray}
Let $z$ be the number of neighbors of a site, where $z=2d$. Then in $d=1$,
 $z=2$, and the difference $\Delta_{i_0}$ defined in Eq. (\ref{deltai}) would be always positive, i.e., the SDRG
would be exact. Alas   it fails to be exact in
higher dimensions as $\Delta_{i0}$ is not always positive.

Monthus argued that ``the worst is not always true.'' Indeed, in a
frustrated spin glass, the worst case discussed above where all the
spins $S_j$ are such that $( J_{ij} S_j)$ have all the same sign, is
atypical.  It is much more natural to compare with a sum of random terms
of absolute values $J_{ij}$ and of random signs, i.e., to replace the
difference $\Delta_i$ of Eq.~(\ref{deltai}) by
\begin{eqnarray}
\Omega_i && \equiv \vert J_{i,j_{\rm max}(i)}  \vert 
	- \sqrt{ \sum_{j \ne j_{\rm max}(i)}  \vert J_{ij}  \vert^2  }.
\label{omegaii}
\end{eqnarray}
Note that for the case of $z=2$ neighbors, $\Omega_i$ actually coincides
with $\Delta_i$, so that the exactness discussed above is the same. But
for $z>2$, it is expected that $\Omega_i$ is a better indicator of the
relative dominance of the maximal coupling for the different spins.
Monthus' version of the SDRG procedure was based on the variable
$\Omega_i$.

At each step, the spin-glass Hamiltonian is similar to that of
Eq.~(\ref{eq:ham}). The variable $\Omega_i$ of Eq.~(\ref{omegaii}) is
computed from the couplings $J_{ij}$ connected to $S_i$.  The iterative
renormalization procedure is defined by the following decimation steps.

\smallskip

\noindent (1) Find the spin $i_0$ with the maximal $\Omega_i$, i.e., 
\begin{eqnarray}
 \Omega_{i_0} \equiv \max_{i} (  \Omega_{i}    ). 
\label{omegaimax}
\end{eqnarray}

\smallskip

\noindent (2) The elimination of the spin $S_{i_0}$ proceeds via
Eq.~(\ref{elimsi0}) and all its couplings $J_{i_0,j} $ with $j \ne
j_{\rm max}(i_0)$ are transferred to the spin $S_{j_{\rm max}(i_0)}$ via
the renormalization rule of Eq.~(\ref{jr}).

\smallskip

\noindent (3) The procedure ends when only a single spin $S_{\rm last}$
is left. The two values $S_{\rm last}=\pm 1$ label the two ground states
related by a global flip of all the spins.

\smallskip

\noindent From the choice $S_{\rm last}=+1$, one can reconstruct all the
values of the decimated spins via the rule of Eq.~(\ref{elimsi0}).

\begin{figure}[htb]
\begin{center}
\includegraphics[width=\columnwidth]{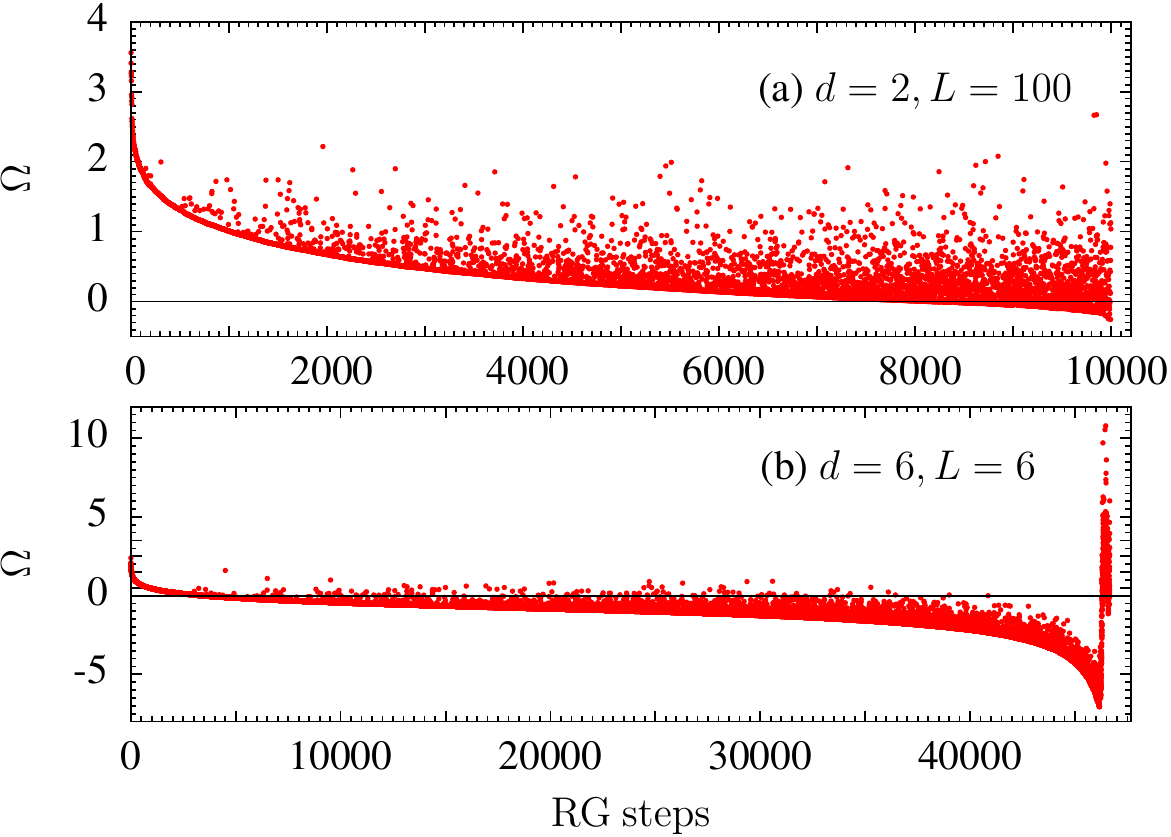}
\caption{
Representative evolution of $\Omega_i$ of the decimated spin as a
function of the RG step, which corresponds to the number of spins which
have been decimated for the EA model for (a) $d=2$ and (b) $d=6$.  Over
most of the iteration range for $d=2$, $\Omega_i$ is positive.  The SDRG
estimate for the exponent $d_{\rm s}$ is also quite accurate in this
case. As $d$ increases, the values of $\Omega_i$ turn negative after a
decreasing number of iterations, suggesting that the SDRG becomes less
accurate in higher dimensions, as can be seen  for $d=6$ [panel (b)].
Note the different horizontal scales.}
\label{OmegaEA}
\end{center}
\end{figure}

Monthus \cite{monthus:15} studied how the value of $\Omega_i$ evolves
with each iteration for the EA model for $d=2$ and $d=3$. For the SDRG
to be exact one needs $\Delta_i$ to be always positive and hopefully 
$\Omega_i$ acts as a useful proxy for $\Delta_i$. She found that
for the early iterations the $\Omega_i$ were indeed positive but turned
negative for the later stages of the iteration procedure, indicating
that the SDRG was failing. She suggested that the fractal dimension
$d_{\rm s}$ was dominated by the early stages of the iteration, which
correspond to long length scales. We have extended her studies of
$\Omega_i$ up to $d=6$ and have found that as the dimension $d$
increases, the crossover where the SDRG would appear to become steadily
worse (i.e., where the $\Omega_i$ turn negative) occurs at successively
earlier stages of the RG iterations. Figure \ref{OmegaEA} shows the form
of the $\Omega_i$ in $d=2$ and $d=6$ space dimensions. Because the SDRG could be exact only if $\Omega_i > 0$ for all $i$, the data for $d=6$ are far from satisfying this criterion.

A defect of the SDRG is that when it terminates it can give a spin state
in which not all the spins are even parallel to their local fields. We
have investigated the problem carefully in two dimensions and found a
small fraction of spins fail to be parallel to their local fields, and
these seem to be the spins which sit in very small values of the local
field. We have generated from these states a one-spin flip stable state
by flipping these spins and their neighbours thereon until there are no
spins left that are not parallel to their local fields. With these new
states we find that the coefficient $A$ in $\Sigma^{DW} \sim A L^{d_s}$
is slightly modified: Its logarithm $\ln(\Gamma)$ is shifted by a small
amount (of order $0.005$) for a wide range of $L$ values.  Because it
does not seem to significantly influence the value of $d_s$, we choose
not to investigate this problem in greater detail here.

\begin{figure}[htb]
\begin{center}
\includegraphics[width=\columnwidth]{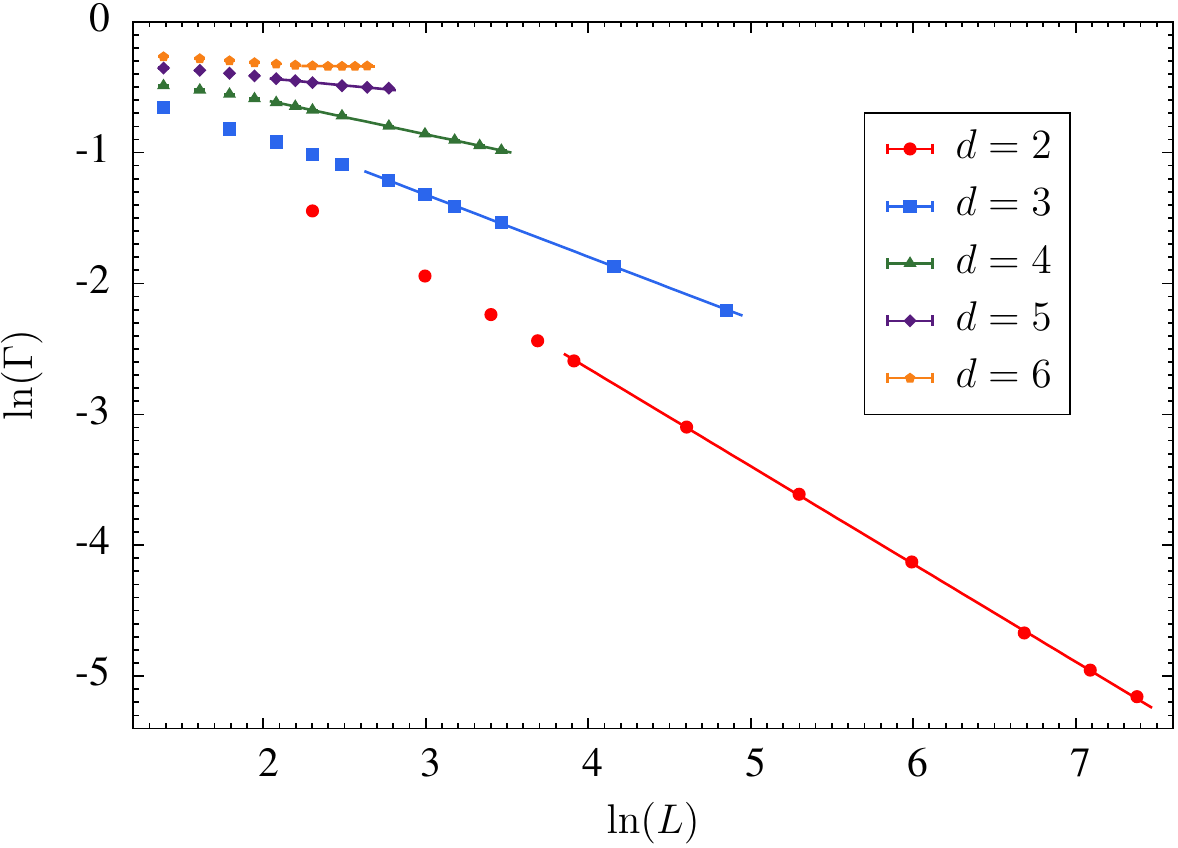}
\caption{
$\ln \Gamma$ for various space dimensions $d$ as a function of $\ln L$
computed using the SDRG algorithm. Note that $\Gamma \sim L^{d_{\rm
s}-d}$. Our estimate of $d_{\rm s}$ is determined by the slope of the
straight lines drawn through the points at large-$L$ values. Note how
the data for $d = 6$ level off, i.e., $d_{\rm s} \to d$. (See Fig. \ref{d6fig}
for an enlarged figure in six dimensions). Error bars are
smaller than the symbols.}
\label{SDRGfig}
\end{center}
\end{figure}

\begin{figure}[htb]
\begin{center}
\includegraphics[width=\columnwidth]{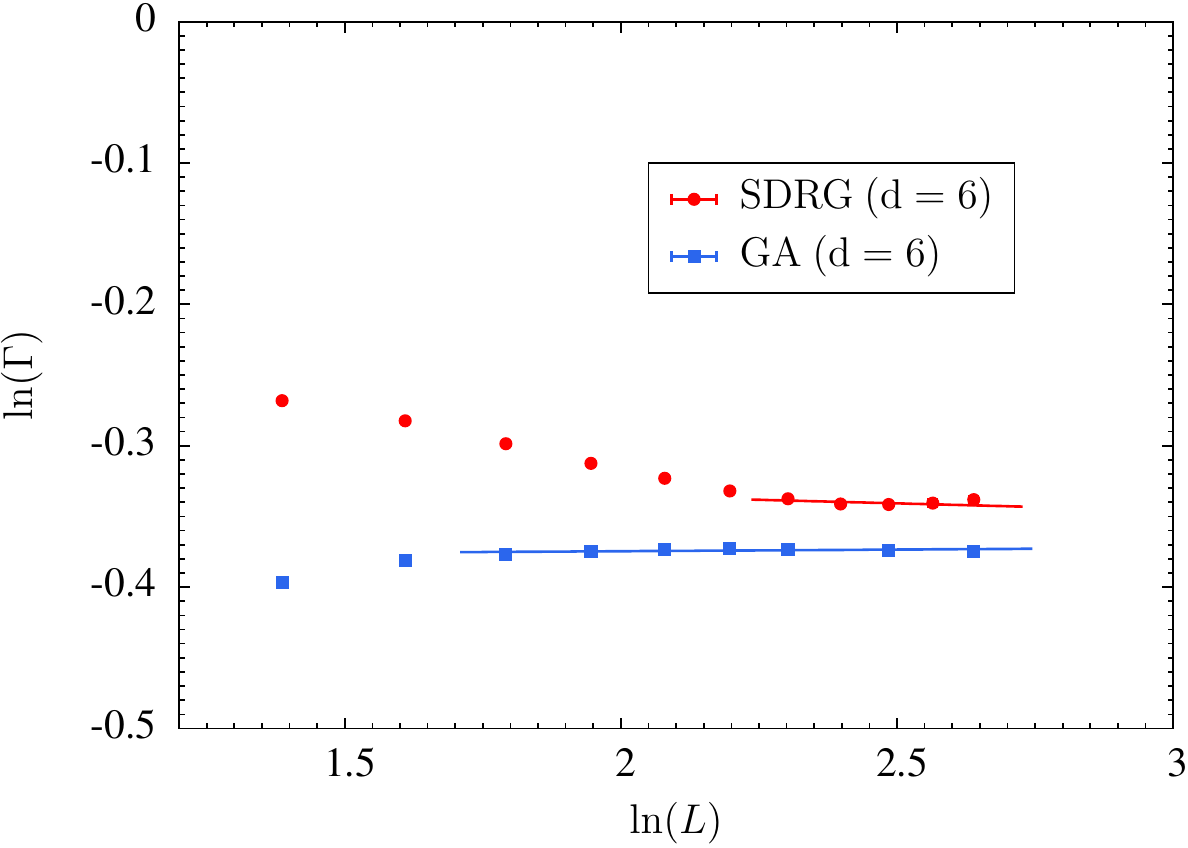}
\caption{
$\ln \Gamma$ for $d=6$ as a function of $\ln L$
computed using the SDRG and GA algorithms. Our estimate of $d_{\rm s}$ is determined by the slope of the
straight lines drawn through the points at large-$L$ values. Using $\Gamma \sim L^{d_{\rm
s}-d}$ the levelling off of the lines at the larger values of $L$ implies that
 $d_{\rm s} \to d$ in six dimensions. Error bars are
smaller than the symbols.}
\label{d6fig}
\end{center}
\end{figure}

\begin{table}
\caption{
Dimensionality $d$, system size $L$, and the number of disorder
realizations $M$ studied using the GA and SDRG methods. Part of the SDRG
data used here are taken from Ref.~\cite{wang:17a}.
\label{table}
}
\begin{tabular*}{\columnwidth}{@{\extracolsep{\fill}} c c c c}
\hline
\hline
Method &$d$ &$L$ &$M$ \\
\hline
SDRG &$2$ &$\{10, 20, 30, 40, 50, 100, 200, 400, 800\}$ &$10000$ \\
SDRG &$2$ &$1200$ &$3000$ \\
SDRG &$2$ &$1600$ &$1000$ \\
SDRG &$3$ &$\{4, 6, 8, 10, 12, 16, 20, 24, 32\}$ &$3000$ \\
SDRG &$3$ &$\{64, 128\}$ &$1000$ \\
SDRG &$4$ &$\{4, 5, 6, 7, 8, 9, 10, 12, 16, 20, 24\}$ &$3000$ \\
SDRG &$4$ &$28$ &$717$ \\
SDRG &$4$ &$32$ &$121$ \\
SDRG &$5$ &$\{4, 5, 6, 7, 8, 9, 10, 12, 14\}$ &$3000$ \\
SDRG &$5$ &$16$ &$1000$ \\
SDRG &$6$ &$\{4, 5, 6, 7, 8\}$ &$3000$ \\
SDRG &$6$ &$9$ &$1843$ \\
SDRG &$6$ &$10, 11, 12$ &$1000$ \\
SDRG &$6$ &$\{13, 14\}$ &$200$ \\[2mm]
GA &$2$ &$\{4, 8, 12, 16, 32, 64, 128, 256, 512\}$ &$3000$ \\
GA &$3$ &$\{4, 6, 8, 10, 12, 16, 20, 24, 32, 64\}$ &$3000$ \\
GA &$4$ &$\{4, 6, 8, 10, 12, 16, 20, 24, 32\}$ &$3000$ \\
GA &$5$ &$\{4, 5, 6, 7, 8, 9, 10\}$ &$6000$ \\
GA &$5$ &$\{12, 14, 16\}$ &$3000$ \\
GA &$6$ &$\{4, 5, 6, 7, 8, 9, 10\}$ &$3000$ \\
GA &$6$ &$\{12, 14\}$ &$1000$ \\
\hline
\hline
\end{tabular*}
\end{table}

\section{SDRG results}
\label{sec:SDRGresults}

In Fig.~\ref{SDRGfig} we plot $\ln \Gamma$ versus $\ln L$ using the SDRG
method of Monthus \cite{monthus:15} to compute the link overlap.  One
change from our previous work in Ref.~\cite{wang:17a} is that we have
added more data. Especially for $d= 6$ we have increased the largest
system studied from $L=10$ to $L=14$. The new data show that for $d=6$
the curve is levelling off, implying that $d_{\rm s} \to d$. We have
also increased the values of $L$ studied in $d=2$ and $3$, going far
beyond the system sizes studied in Ref.~\cite{monthus:15}.
Table~\ref{table} lists simulation parameters, such as the number of
bond configurations $M$ for each value of the linear system size $L$ in
space dimension $d$.

The SDRG seems to give quite accurate results for the value of $d_{\rm
s}$ at least in low space dimensions. Thus, in $d = 2$, Monthus found
from the SDRG a value of $d_{\rm s} \approx 1.27$ from $L$ values up to
$340$, a result which is similar to a recent study of systems up to $L =
10^4$ \cite{khoshbakht:17a}  based on fast polynomial time algorithms for
finding ground states (which, however, only work in two space
dimensions) which gives $d_{\rm s} =1.27 319(9)$. In $d=3$, Monthus
finds $d_{\rm s} =2.55$ for systems of size up to $L=45$. In
Ref.~\cite{wang:17c} a value of $2.57$ is quoted from studies on systems
up to $L=12$. The SDRG is just an algorithm which attempts to find the
ground-state spin configuration.  It is exact in one space dimension.
While it seems to give excellent values for $d_{\rm s}$, it gives poor
values for the actual ground-state energy itself and the energy cost of
the interface.  If the domain-wall energy scales $\sim L^{\theta}$, then
Monthus reports $\theta \approx 0$ whereas the  recent high-precision
calculations  show that $\theta = -0.2793(3)$ \cite{khoshbakht:17a}.

\begin{figure}[htb]
\begin{center}
\includegraphics[width=\columnwidth]{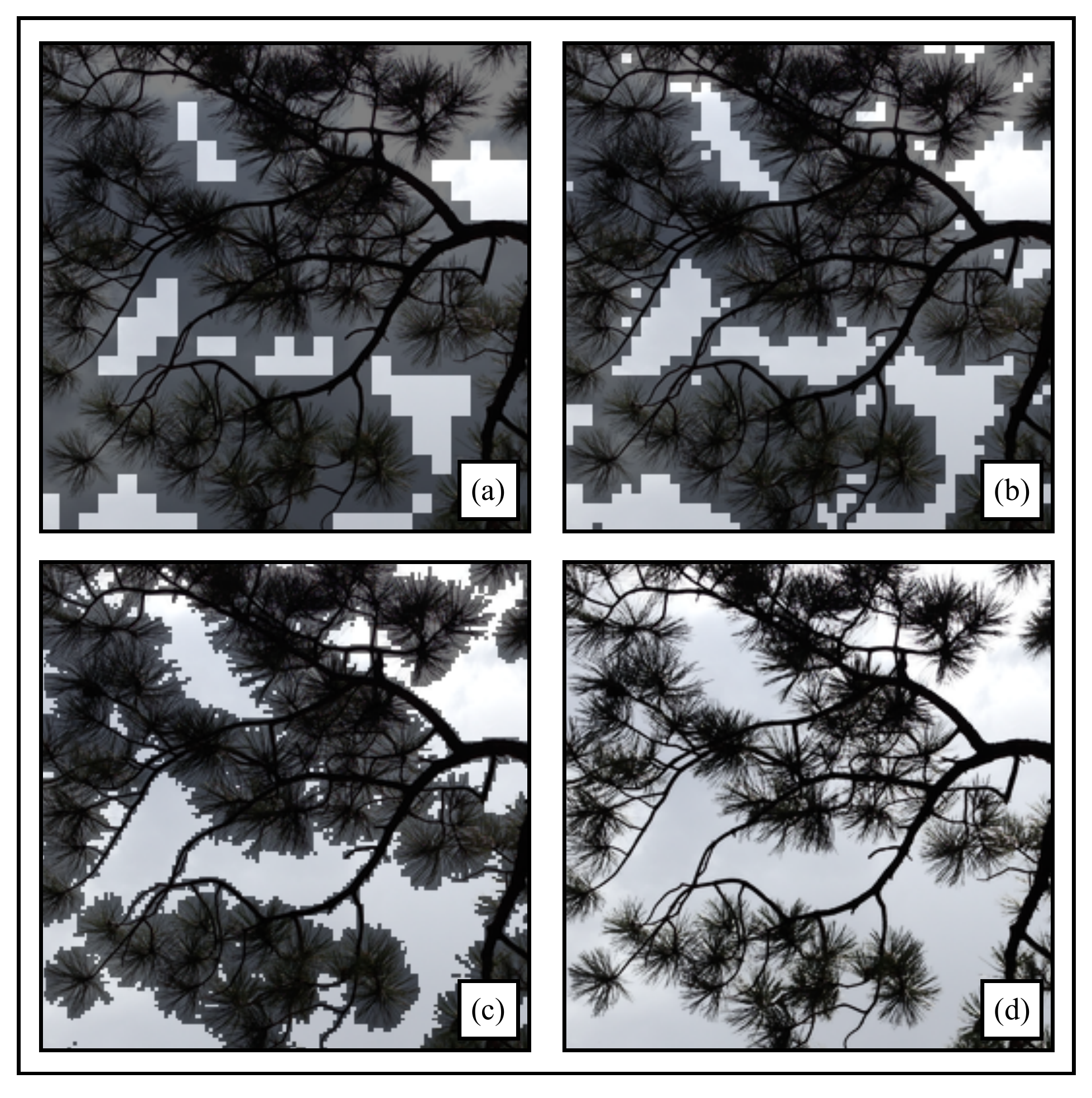}
\caption{
The bifurcation of a tree is a self-similar fractal. The four figures
are measurements of its length using square domains whose linear size is
reduced at each step of the renormalization. For a self-similar
fractal, like the ponderosa pine depicted here, the scaling dimension
$d_{\rm s}$ is the same no matter what length scale is used to
determine it. Panel (a) shows the coarsest measurements which are
successively refined by reducing the size of the squares in panels (b) -- (d). Note that the domains
are smaller than the image resolution in panel (d). The fractal
dimension of the ponderosa pine is approximately $1.88$. One could in principle
obtain the correct fractal dimension by studies at the coarsest length scales which is why we suspect that the SDRG, which works better on the coarsest length scales, is capable of getting accurate answers for $d_{\rm s}$. }
\label{selfsimilar}
\end{center}
\end{figure}

Because Monthus' value for $d_{\rm s}$ in $d=2$ seemed to be compatible
with the high-precision calculations  \cite{khoshbakht:17a}, we
speculated in Ref.~\cite{wang:17a} that the SDRG might be accurate
because the interface is a self-similar fractal \cite{mandelbrot:67}.
The SDRG seems to be accurate in the early stages of the RG process
where the $\Omega_i$ are positive (see Fig.~\ref{OmegaEA}) where a
coarse approximation of the domain lengths is performed (see
Fig.~\ref{selfsimilar}).  In the later stages of determining  the domain
length, the SDRG's accuracy will decrease. In particular, in the
relation $\Sigma^{DW} \sim A L^{d_{\rm s}}$ we suspect that the SDRG
might determine $d_{\rm s}$ quite accurately, but that the coefficient
$A$ might be obtained with less accuracy.  To estimate $A$ to high
accuracy would require an RG process accurate on all length scales, both
short and long.  In this paper we have extended the system sizes studied
far beyond those studied by Monthus in $d=2$, and find that $d_{\rm
s}=1.2529(14)$ which indicates that the SDRG is not exact for $d_{\rm
s}$ in $d=2$, but just a good approximation. Our estimate of $A$ is
$1.4040(106)$ whereas the recent high-precision estimate is $1.222(3)$
\cite{khoshbakht:17a}.

We have also extended Monthus' work in $d = 3$ from $L = 45$ to $L= 128$
and find $d_{\rm s} = 2.5256(30)$. If we had only system sizes up to
$12$ in $d=3$, as in the Monte Carlo studies of Ref.~\cite{wang:17c},
then because of finite-size effects (visible in Fig.~\ref{SDRGfig}), we
would have reported a value of $d_{\rm s} \approx 2.6093(50)$.  A value
of  $2.57$ was reported in Ref.~\cite{wang:17c}
based on the same range of $L$ values up to $L=12$.

The SDRG is not an analytical treatment, but a numerical technique and
in high dimensions (e.g., $d = 5$ and $6$) this limits us to studying
rather small linear system sizes. As a consequence, estimates of
exponents can be affected by finite-size corrections as aforementioned
for $d=3$. Thus, it is hard to be certain that $d_{\rm s}=d$ in six
dimensions. We therefore decided to also use a greedy algorithm (GA) to
complement the SDRG results. It is already known from analytical studies
that $6$ is the ''upper critical dimension''  for the  GA, at least for the fractal dimension associated with minimum spanning trees.
\cite{jackson:10,jackson:10a}.  Here, we want to know whether numerical
studies of the value of $d_{\rm s}$  would also show that
six is a similarly special dimension for the fractal dimension of domain walls with the GA algorithm.

\begin{figure}[htb]
\begin{center}
\includegraphics[width=\columnwidth]{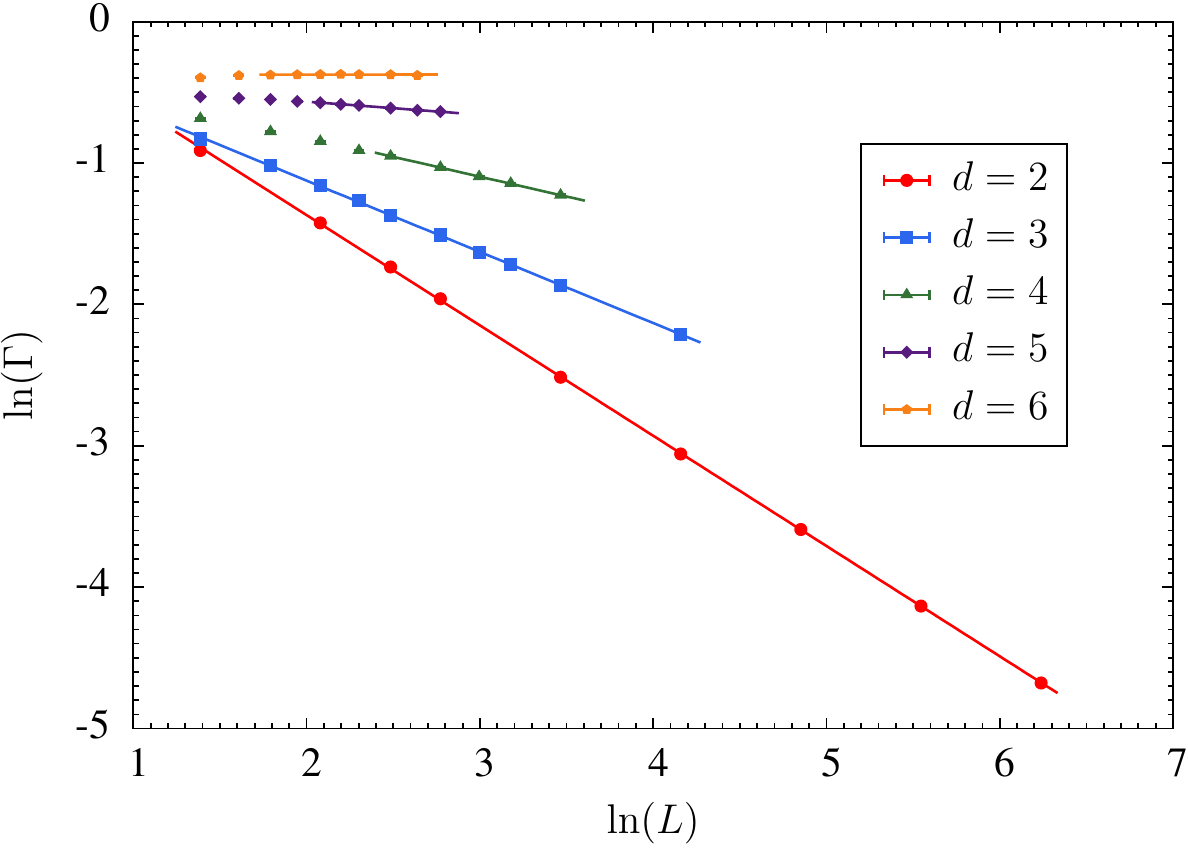}
\caption{
$\ln \Gamma$ for various space dimensions $d$ for the EA model as a
function of $\ln L$ computed using the GA. Note that $\Gamma \sim
L^{d_{\rm s}-d}$. Our estimate of $d_{\rm s}$ is determined by the slope
of the straight lines drawn through the points at large $L$-values.
Error bars are smaller than the symbols.}
\label{GAfig}
\end{center}
\end{figure}

\section{The greedy algorithm}
\label{sec:greedy}

The GA (also studied by Monthus \cite{monthus:15}) works as follows. The
bonds in the order of decreasing absolute magnitude are satisfied in
turn, unless a closed loop appears then the bond is skipped, until the
relative orientation of all the spins is determined.  In Table
\ref{table}, we have given details of the system sizes and numbers of
different bond realizations which we have studied in dimensions $d=2$,
$\cdots$, $6$. In Fig.~\ref{GAfig} we plot $\ln \Gamma$ versus $\ln L$
determining the link overlap using the GA. Notice that the corrections
to scaling in $d = 6$ seem smaller for the GA than for the SDRG method,
because the data seem independent of $L$ even for the smallest system
sizes.

\begin{figure}[htb]
\begin{center}
\includegraphics[width=\columnwidth]{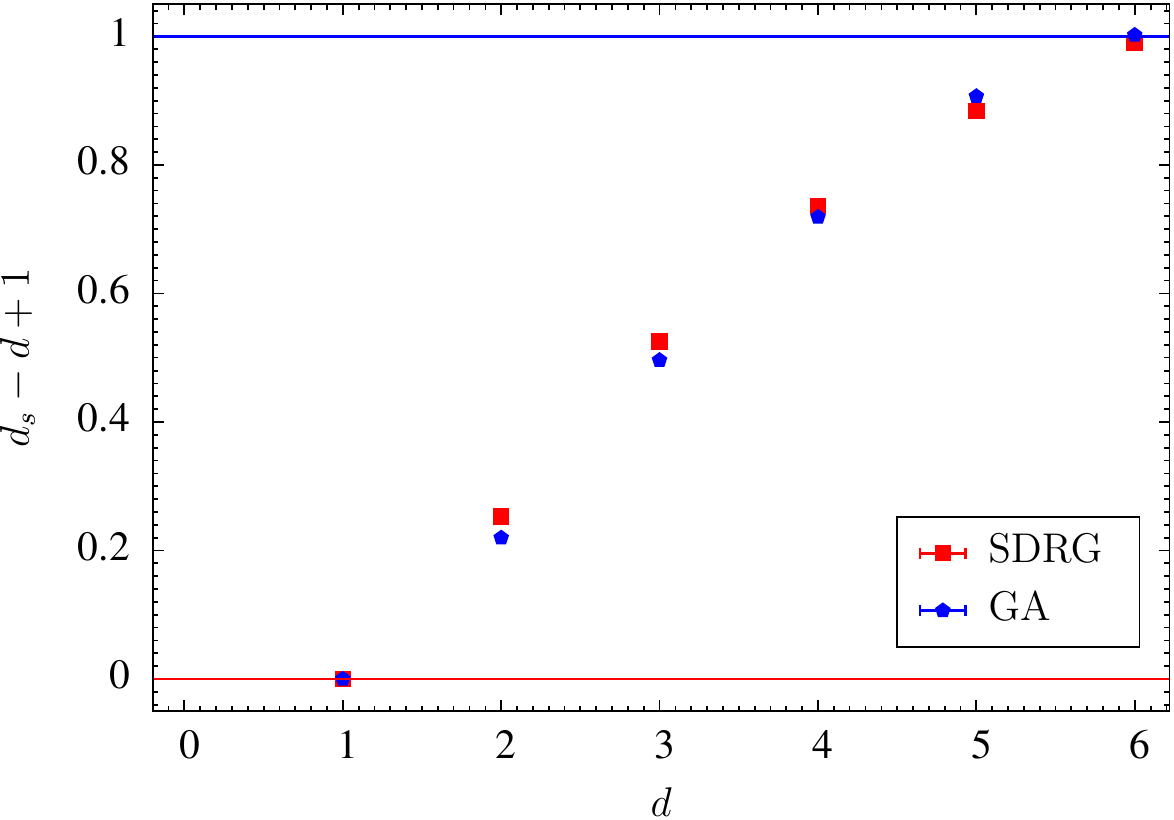}
\caption{
Greedy algorithm (GA) results (blue pentagons) compared with
strong-disorder renormalization group (SDRG) results (red squares) for
$d =2$, $3$, $4$, $5$, and $6$. The upper bound $d_{\rm s}-d+1$ at unity is
marked by a horizontal blue line, while the lower bound at zero is
marked with a horizontal red line. The value $d_{\rm s}=0$ for $d=1$ is
exact and given by both methods. Only statistical errors are included
and error bars are smaller than the symbols. Numerical values are summarized in Table.~\ref{table2}.
}
\label{GA3}
\end{center}
\end{figure}

Like the SDRG procedure, the GA is just a way of finding the spin
configuration for a putative ground state of the system. There is no
bond renormalization as in the SDRG [see Eq.~(\ref{jr})]. It is just as
poor for the ground-state energy and the exponent $\theta$ as the SDRG
\cite{monthus:15}. In $d =2$ we obtain $d_{\rm s}^{\rm GA} \simeq
1.2196(11)$, which is comparable with 
Ref.~\cite{sweeney:13} who quote $d_{\rm s}^{\rm GA}=1.216(1)$.  Note
that the SDRG value for $d_{\rm s}$ is in much better agreement with the
high-precision value of Ref.~\cite{khoshbakht:17a}.  In $d =3$ the GA
result is $d_{\rm s}^{\rm GA} \simeq 2.4962(19)$, which is closer to
that of the SDRG. An earlier estimate in three dimensions is that of
Ref.~\cite{cieplak:94} who quote $d_{\rm s}^{\rm GA} \simeq 2.5 \pm
0.05$. In Fig.~\ref{GA3} we have plotted $d_{\rm s}-d+1$ versus $d$
using the $d_{\rm s}$ from both the GA and SDRG algorithms. As the
dimension $d$ approaches $6$ the two estimates appear to merge and give
$d_{\rm s}=d$ in $d = 6$. The analytical expectation  of
Refs.~\cite{jackson:10,jackson:10a} was that $6$ is the upper critical dimension for the fractal dimension of minimum spanning trees  within the GA.  Our numerical work suggests that within the GA, domain walls also have $6$ as their upper critical dimension.

\section{Discussion}
\label{sec:discussion}

\begin{table}
\caption{
Numerical estimates of the fractal dimension $d_s$ of the SDRG and GA methods. $d_s$=0 for $d=1$, as both methods are exact for the one-dimensional model. Error bars are statistical errors.
\label{table2}
}
\begin{tabular*}{\columnwidth}{@{\extracolsep{\fill}} c c c c c c c}
\hline
\hline
Method &$d=2$ &$d=3$ &$d=4$ &$d=5$ &$d=6$ \\
\hline
SDRG &1.2529(14) &2.5256(30) &3.7358(36) &4.884(60) &5.9899(60) \\
GA &1.2196(11) &2.4962(19) &3.7190(47) &4.9068(32) &6.0023(22) \\
\hline
\hline
\end{tabular*}
\end{table}

We have obtained numerical results (Fig.~\ref{GA3}) using 
a strong-disorder renormalization group method and a greedy algorithm
that are consistent with $6$ being a special space dimension above
which the conventional EA model with a Gaussian bond distribution has
RSB behavior and summarized them in Table \ref{table2}.  For $d \le 6$, we have found that within our numerical procedures that the EA model is  behaving according to droplet model expectations because $d_{\rm s} < d$. That $6$ is a special dimension for the behavior of spin glasses is  in
accord with some older expectations based on analytical results
\cite{bray:80,moore:11}, but these  have been controversial \cite{parisi:12,moore:18x}.  Because both the GA and the SDRG are
approximations, we regard the results presented here as not decisive.

We note, however, that real-space RG methods such as the SDRG are
capable of endless refinements. Monthus \cite{monthus:15} herself
discussed a variant, the ``box'' method, which improved the value of the
zero-temperature exponent $\theta$ in $d=2$ from the very poor value
$\theta \approx 0$ obtained by the SDRG method described in this paper
to at least a negative value of $\theta \approx -0.09$ [the
high-precision estimate of Ref.~\cite{khoshbakht:17a} is
$\theta=-0.2793(3)$]; note that the value of $d_{\rm s}$ was hardly
altered.  It might be possible to find a real-space RG procedure that
gives accurate numbers on all quantities of interest for
three-dimensional spin glasses. The SDRG and the GA have a common
feature in that they both recognize that the largest bonds are likely to
be satisfied in the ground state. We suspect that will be an ingredient
of any future successful RG scheme for spin-glass systems.

\begin{acknowledgments}

M.A.M.~would like to thank Nick Read for email discussions. We thank
Martin Weigel for supplying more details of his results.  H.G.K.~would
like to thank Della Vigil at the Santa Fe Institute for helping with
determining the type of pine photographed in Fig.~\ref{selfsimilar} and
appreciates award No.~06210311-251521-23011407. W.W.~acknowledges
support from the Swedish Research Council Grant No.~642-2013-7837 and
Goran Gustafsson Foundation for Research in Natural Sciences and
Medicine.  W.W.~and H.G.K.~acknowledge support from NSF DMR Grant
No.~1151387. The work of H.G.K.~and W.W~is supported in part by the
Office of the Director of National Intelligence (ODNI), Intelligence
Advanced Research Projects Activity (IARPA), via MIT Lincoln Laboratory
Air Force Contract No.~FA8721-05-C-0002. The views and conclusions
contained herein are those of the authors and should not be interpreted
as necessarily representing the official policies or endorsements,
either expressed or implied, of ODNI, IARPA, or the U.S.~Government. The
U.S.~Government is authorized to reproduce and distribute reprints for
Governmental purpose notwithstanding any copyright annotation thereon.
We thank Texas A\&M University for access to their Ada and Curie
clusters.

\end{acknowledgments}

\bibliography{refs}

\end{document}